\documentstyle[preprint,revtex,eqsecnum]{aps}

\begin{document}
\draft
\begin{title}
\begin{center}
Level repulsion in integrable and almost-integrable\\
quantum spin models\end{center}
\end{title}
\author{Theodore C. Hsu\cite{byline} and J.C. Angl\`es d'Auriac\cite{jbyline}}
\begin{instit}
Centre des Recherches sur les Tr\`es Basses Temp\'eratures,\\
BP 166, 38042 Grenoble, France
\end{instit}
\receipt{}
\begin{abstract}
The repartition of the separation between energy
levels of various isotropic S=1/2 antiferromagnetic chains
is studied numerically with the aim of investigating
the transition from integrable to non-integrable systems.
We begin by displaying the level separation distribution
of the integrable Bethe chain. Then two non-integrable
systems, two coupled chains and a next-nearest-neighbor
coupled chain, are studied as a function of the
coupling. We examine how the level spacing
evolves from the Poisson distribution to the GOE distribution.
Finally we consider
the Haldane-Shastry $1/r^{2}$ model.
A number of conclusions regarding the behaviour
and relevance of the level spacing distribution in
these spin systems is drawn.
\end{abstract}
\pacs{PACS numbers:  75.10.Jm, 05.45.+b}
\narrowtext

\section{Introduction}
Interest in strongly interacting
Fermion systems has recently been invigorated with the discovery of
high temperature superconductors.
Another strongly interacting Fermion system is the atomic nucleus.
The stability of nuclei and description of their low-energy excitations
have been understood
for many years. But the higher energy excited states of the nucleus
are complex and can only be described statistically.
Wigner \cite{WIGNER} suggested that the Hamiltonian
of this system should be similar to a random matrix
and that the distribution of spacings of nuclear energy
levels should reflect this \cite{PORTER,BOHIGAS}.
In particular the Gaussian Orthogonal Ensemble (GOE)
of N$\times$N real symmetric matrices, invariant under
orthogonal transformations, with random matrix elements
that are Gaussian distributed (zero-mean, variance $v^{2}$
[diagonal elements have variance $2v^{2}$])
has the following properties:

a) The ensemble-averaged density of states  has the
	elliptical form \cite{WIGNERB}
	$\rho(x) = \sqrt{4 - x^{2}}/2\pi$, where
	$x = E/\sqrt{Nv^2}$, $|x|<2$ and zero
	otherwise. This is referred to as Wigner's
	semicircle law.

b) The probability that the eigenvalues
	be $\lambda_{1},\cdots,\lambda_{N}$
	is \cite{WISHART},
\begin{equation}
P(\lambda_{1},...,\lambda_{N})
=
{
{
2^{N(N-1)/4}
}
\over
{
n!(2v)^{N(N+1)/2}
\prod_{g=1}^{N}\Gamma({g\over 2})
}
}
e^{
-\sum_{i}\lambda_{i}^{2}/4v^{2}
}
\prod_{i,j}|\lambda_{i}-\lambda_{j}| \quad .
\end{equation}

The last term in the above equation gives rise to
energy level repulsion. The distribution of spacings between
pairs of energy levels has been found empirically
to be quite accurately described by the `Wigner surmise'
\cite{WIGNER}
based on two dimensional matrices. This `surmise' is that
the probability that the spacing between two adjacent
levels be $s$ is $P(s) = (s\pi/2)\exp{(-s^{2}\pi/4)}$ where
the probability has been normalized so that $\langle s\rangle = 1$.

In contrast integrable systems, which have as many constants
of the motion as degrees of freedom, and for which each energy
level can be labelled by that many quantum numbers, have
generically a Poisson distribution, $P(s) = e^{-s}$,
for the energy level spacing \cite{BERRY}. The Hamiltonians
of these systems can
be thought of being representable by random {\it diagonal} matrices.
The interpolation between Poisson and GOE distributions has been
modelled by so-called `band random matrix ensembles' (BRME)
\cite{CASATI}.
In these ensembles only the off-diagonal elements
in some sense `close' to the diagonal are non-zero and
random. This is meant to interpolate between the random diagonal
matrix and the GOE in which all off-diagonal elements are non-zero and
random.
A BRME might be relevant for local tight-binding models
since in a natural basis of states
(e.g. the one that is diagonal in $S^{z}_{i}$, for all sites i)
only a few entries will have a non-zero value.
On the other hand the non-zero matrix elements will
be scattered about and not all close to the diagonal.
Moreover expressing the Hamiltonian in a basis
where all the obvious symmetries
are also diagonal will
leave us with a block-diagonal matrix where all the blocks will have
only non-zero entries
(particularly when we diagonalize with respect to total spin).
Therefore
we do not see how to justify using the BRME to interpret
our results.

The theoretical motivation for the studies undertaken here
and by Montambaux {\it et al.} \cite{MONT} is
the search for a microscopic theory of high temperature
superconductors. In the `normal' state of these materials
they are not Landau-Fermi liquids \cite{PWA}. One of
the challenges in the field is to prove or disprove
the existence of a Fermi liquid in
two-dimensional interacting electron models.
In a Landau-Fermi liquid
the momenta and spin
$\{k_{i},\sigma_{i}\}$ of excited quasiparticles
form a set of good quantum numbers.
Those who
have modelled high temperature superconductors
by a strongly interacting
model such as the large-U Hubbard model or t-J model and
tried to construct elementary single-particle excitations
have not had success in constructing elementary
quasiparticle excitations which are weakly interacting.

In one dimension, Fermi liquid behaviour, and the
set of momenta, charge and spin of Landau quasiparticles
may be replaced by the integer parameters of a Bethe Ansatz
solution of an interacting model (if one exists).
If this does not happen in two dimensions, how then
might Landau-Fermi liquid behaviour disappear?
It has been proposed \cite{RAMMAL} that perhaps there
{\it do not exist} weakly interacting quasiparticles whose
momenta, charge and spin would be a `good' set
of quantum numbers. In that case, the absence of `good' quantum
numbers might be signalled by level statistics resembling those
of random matrices. If, on the other hand, an interacting
Fermion system retained Landau-Fermi liquid
behaviour one might expect to see the level statistics
of an integrable system, especially for low energy excitations.
The first numerical study along these lines was performed by
Montambaux {\it et al.} \cite{MONT} who showed
that a special case of the
doped t-J model has a level distribution agreeing quite well with
that of the GOE.

In order to investigate the transition between integrability,
and non-integrability
we have studied the energy level
spacing in two integrable quantum spin systems
and related, but non-integrable,
models which may be obtained from an integrable one by
tuning a
single parameter. The primary integrable model
we worked with was the S=1/2 antiferromagetic chain.
This model is well studied and enables us to compare
the behaviour of the level separation distribution
with the known properties of this system as it is perturbed.

\section{Numerical procedure}
For simplicity and clarity we chose to work
with isotropic spin systems. In this case the total spin and
total $S^{z}$ are good quantum numbers.
It is only necessary to consider the subspace
$S^{z}=0$ which contains all of the eigenenergies.
Open boundary conditions were chosen.
The eigenstates
are representations of the trivial spatial
symmetries, namely, reflection and rotation in spin space.
Thus they can be grouped according to their respective
quantum numbers, parity P and total spin S.
The perturbations which carry
the system from an integrable to a non-integrable one
will always respect the trivial spatial symmetries.
States in different (P,S) sectors will never be
coupled and their energy levels never correlated.
Thus we calculate the energy level separation distribution
within each (P,S) sector separately. In order to
obtain good statistics one requires a large number
of states in each (P,S) sector. It was for this reason
that periodic boundary conditions for the spin chain
were not used. In that case the parity under reflection P
(which takes the values $\pm 1$),
would be replaced by momentum which takes on L values where
L is the length of the chain.

The Hamiltonian was diagonalized numerically using the Jacobi
method. States were sorted by energy, S and P.
In order to correct for gross variations
of the density of states as a function of energy
the level spacing was normalized by the smoothed
local density of states.
This process is commonly referred to as unfolding
the spectrum to remove a fluctuating local level density.
This was performed
for each set of quantum numbers separately.
In general this did not affect the level spacing distribution
very much at all as the density of states was generally constant
with a fall-off at the `band' edges.
The states with energies near the band edges were
discarded by dividing the states of each (P,S) sector into
ten bins ordered by energy and of equal width. The states
of the first and last bins were then discarded.
In order to compare with the statistical
distributions, the energy level separations were
normalized to have a mean of unity. After discarding
results from (P,S) sectors with less than 50
states (generally the high spin sectors) the results
of remaining sectors were combined  at this point
point in order to improve the statistics.
We checked that all of the sectors had roughly the
same behaviour. The probability
function $P(s)$ was plotted by binning the data and
again normalizing the number of states in each bin
so that $\int P(s) ds = 1$.
In some cases $I(y) = \int_{0}^{y}P(s)ds$ was calculated so this
last binning step could be skipped.

It is useful to know, given the typical
size of the Hamiltonian
matrix (after sorting by the straightforward symmetries),
how well the level spacings of the
eigenvalues of a random matrix of that size follow
the Wigner surmise. The linear matrix
dimensions encountered in this work are
in the range 100 -- 500.
Although this sort of calculation has been published
in the past \cite{PORTER},
we have re-done such a calculation
and in Fig. \ref{RMT} we show the
$P(s)$ obtained (by the method described above)
with the matrices of
the same size as encountered in
the (P,S) sectors of an open L=14 site chain.
Fig. \ref{RMT} shows the level of `noise' expected even
if the random matrix hypothesis is satisfied. Also shown
is the `noise' level expected upon comparing
$P(s)$ of random diagonal matrices of these sizes to a Poisson
distribution.

\section{Bethe Chain}
We begin by examining the familiar S=1/2 antiferromagnetic chain
on L sites with nearest-neighbor coupling,
\begin{equation}
H = \sum_{i=1}^{L-1}
J{\bf S}_{i}\cdot{\bf S}_{i+1}
\quad .
\end{equation}
This model remains integrable with our choice
of open boundary conditions.
In Fig. \ref{BETHE} we show P(s) for chains
of length L=12 and L=14 (L=13 is similar but for clarity
it is not shown). They are compared with
the Poisson distribution and Wigner's surmise for the GOE.
Consistent with the integrability of this system
the agreement with the Poisson distribution is
good, especially in the tail. There is a small deviation
from the Poisson distribution at intermediate $s$ which
is more than that of random diagonal matrices
of similar dimensions.
The comparison of L=12 and L=14 cases
gives an idea of finite size effects which may explain this
deviation.

\section{Two coupled chains}
In this section we consider two open
chains. They are coupled by a simple nearest-neighbor
interaction,
\begin{equation}
H =
\sum_{i=1}^{L-1}
\sum_{j=1}^{2}
J{\bf S}_{i,j}\cdot{\bf S}_{i+1,j}
+
\sum_{i=1}^{L}
J_{\perp}{\bf S}_{i,1}\cdot{\bf S}_{i,2}
\end{equation}
where $j=1,2$ is the chain index.
An additional symmetry, reflection between chains,
appears and so the eigenstates were also sorted
by parity under this reflection.
For zero coupling the system is integrable and
when coupling is turned on the system is not integrable
(in fact our calculation erases any doubt that this system
might have been integrable).
Such a coupling between two chains
is believed to be a relevant perturbation
for the ground state \cite{SAKAI}.
We studied this system with the hope that perhaps
the relevance or irrelevance of the inter-chain coupling
might be apparent in the level spacing distribution.
That is, if the coupling were irrelevant, the spectrum
would look like that of two integrable chains. If the
coupling were relevant than the spectrum would look
like that of an non-integrable system.

The distribution $P(s)$ for two chains
of length L=7 is plotted in Fig. \ref{JPERP}
for various values of $J_{\perp}/J$. There is an evolution
from a Poisson distribution to the Wigner surmise as
$J_{\perp}/J$ is turned on. In order to quantify
the evolution between these two distributions we shall
describe it by the single parameter \cite{SHKLOVSKII}
$I = \int_{0}^{\eta} P(s)ds$ where $\eta\approx 2.002$
is the greater of the two values of $s$ where the Poisson
and GOE distributions cross. At the crossing point $I$
is most sensitive to the difference between Poisson and
GOE distributions. For the Poisson and GOE distributions
$I$ has the values 0.8649 and 0.9571 respectively.

In Fig. \ref{LOGJPERP}, $I$ is plotted as a function
of ${\rm ln}(J_{\perp}/Jd)$ separately for
the values S=0, 2, 4, 6, 8 of total spin.
The parameter $Jd$ is the average
spacing between energy levels for a given value of S.
It was extracted from our numerical results.
We subtracted the (empirical) small $J_{\perp}$
limit of $I$ before plotting because $I$ did not
converge to the ideal Poisson distribution value of 0.8649
(probably due to finite size effects).
Fig. \ref{LOGJPERP} shows
that the transition from Poisson to GOE is roughly the
same in the different spin sectors averaged to arrive
at Fig. \ref{JPERP}.
Our results are consistent with the idea that
in general, level repulsion will be
fully developed when the typical
energy shift due to a perturbation
is of order the typical spacing between unperturbed
energy levels. For two chains of length L=7,
the average level spacing of the large
sectors ranges from 0.03 -- 0.07 J. The expectation
value of the $J_{\perp}$ perturbation
is difficult to estimate but there are
seven links between the two chains and
the rough order of magnitude of
$\langle{\bf S}_{i,1}\cdot{\bf S}_{i,2}\rangle$
should be $1/4$.
Thus before ${\rm ln}(J_{\perp}/Jd)$
reaches $-1$ or so level repulsion should have set in.
This is observed and so our results are consistent
with a transition to non-integrability for arbitrarily
small $J_{\perp}/J$ in the thermodynamic limit.
Another fact supporting the idea of comparing
mean energy spacings with the size of the perturbation
is that the
level spacing distributions are roughly similar
if one changes the sign of $J_{\perp}$.

\section{Next-nearest-neighbor-coupled chain}

We now consider a chain with next-nearest-neighbor
(NNN) coupling,
\begin{equation}
H =
\sum_{i=1}^{L-1}
J{\bf S}_{i}\cdot{\bf S}_{i+1}
+
\sum_{i=1}^{L-2}
J_{2}{\bf S}_{i}\cdot{\bf S}_{i+2}
\quad .
\label{NNN}
\end{equation}
For $J_{2}/J = 0$ this system is of course integrable.
Near $J_{2}/J=0.24$ it is believed that the ground
state of this system undergoes a transition from
a liquid-like to
a dimer-like ground state \cite{JULLIEN}.
At $J_{2}/J=0.5$ the ground
state is known \cite{MAJUMDAR}, and is simply
the (doubly degenerate) dimer solid. It would be
interesting to see if this qualitative
behaviour of the ground state is at all reflected
in the level spacing distribution $P(s)$.
One factor which may be significant is
the proximity of the integrable `$1/r^{2}$'
model \cite{HALDANE}, which is
discussed in the following section.

In Fig. \ref{J2} we plot P(s) for a number of values
of $J_{2}/J$ with L=13.
There is no special behaviour near the point $J_{2}/J = 0.24$
except that
level repulsion settles in
continuously but more slowly (as a function
of $J_{2}$ or $J_{\perp}$)
than for the coupled chain
problem.
To illustrate this explicitly we plot the parameter $I$
versus ${\rm ln}(2J_{2}/Jd)$ in Fig. \ref{LOGJ2}.
Subtracted from $I$ is its empirical value when $J_{2}=0$.
$Jd$ is again the observed average level spacing which
is different for sectors of different total spin.
The extra factor of two
appears because
NNN coupling introduces one
coupling per site whereas in the inter-chain coupling problem
there is one extra coupling for every two sites.
Comparing Fig. \ref{LOGJPERP} to Fig. \ref{LOGJ2}
one sees that the parameter $I$ starts to deviate from
its value for the integrable case
at a larger value of ${\rm ln}(2J_{2}/Jd)$
than ${\rm ln}(J_{\perp}/Jd)$ for coupled chains.
So in terms of
affecting integrability
it would seem that
interchain coupling is a somewhat stronger perturbation
than NNN coupling. This behaviour is also evident
upon examining the whole integrated probability
distribution curves $I(y) = \int_{0}^{y}P(s)ds$
for these models.

Another way of understanding the above
observation is that the resistance to level
repulsion might be due to the proximity of the integrable
$1/r^{2}$ model Eq. \ref{SR2}.
When $J_{2}/J = 0.25$ the Hamiltonian
\ref{NNN} contains the first two terms
of Eq. \ref{SR2}.
In order to test this idea we evaluated the
level spacing distribution for a {\it ferromagnetic}
($J_{2}/J < 0$) coupling. We found that the level
spacing distribution as a function of $|J_{2}/J|$
behaved essentially the same as for antiferromagnetic
NNN coupling. So the proximity
of the $1/r^{2}$ model is perhaps not responsible.
Another, less likely, possibility is that there is a
hitherto unknown integrable model nearby with
ferromagnetic NNN coupling!

\section{$1/r^{2}$ chain}

The spin 1/2 periodic chain with Hamiltonian
\begin{equation}
H = \sum_{i,n} {J\over 2}{\rm sin}^{-2}(n\pi/L)
{\bf S}_{i}\cdot{\bf S}_{i+n}
\label{SR2}
\end{equation}
was studied by Haldane and Shastry \cite{HALDANE}
and shown to be integrable.
We have studied an open chain version,
\begin{equation}
H = \sum_{i,j=1;i\ne j}^{L}
{J\over 2}|i-j|^{-2} {\bf S}_{i}\cdot{\bf S}_{j}
\end{equation}
in order
to avoid the appearance of L conserved momenta
which would reduce the statistical significance
of the level spacing distribution.
The results are summarized in Fig. \ref{USR2}.
The level spacing distribution is strikingly
unusual in that that the
probability of closely spaced levels is {\it larger} than for a
Poisson distribution.
One way that this might arise
is through the Landau levels of an external magnetic field
but no such field is present here. It would be interesting to
know if recent advances in understanding this model \cite{HALDANEA}
could explain this behaviour.

\section{Conclusions}

In this work we have studied the level
spacing distribution for interacting quantum
many-body systems represented by
antiferromagnetic spin 1/2 chains.
We have confirmed that the level spacing
distribution for the integrable Bethe chain is Poissonian
and that certain perturbations lead to level repulsion.
A system was found, the $1/r^2$ model, which
displays level attraction.
We were able to track the transition from integrability
to non-integrability.
We conclude, by a small system diagonalization, that
certain systems such as two coupled chains or
the NNN coupling model (irrespective of the sign
of coupling) are definitely {\it not} integrable.
A possible problem which did not appear was
that of a long chain `almost' having
translation invariance and hence an `almost' good
momentum quantum number. That would introduce additional
degeneracies in the non-integrable models, but none were seen.

Level repulsion seems to set in, as one might guess,
when the perturbation is of the
same size as the typical spacing between
energy levels. In the thermodynamic limit
the extension of this idea would require some care.
One would need to scale both the
energy level spacing and perturbation with system
size. Additional complications would arise were
one to also consider a low energy limit
where the density of states is changing rapidly.

We found evidence that the introduction
of a second space dimension
has a slightly stronger effect on integrability than
the introduction of the NNN coupling.
In the NNN coupling study we saw no
special behaviour near
$J_{2}/J = 0.24$ other than a resistance
to level repulsion. It is perhaps not surprising that
a qualitative change in the ground state does not affect
the level statistics of the bulk of the states.
On the other hand at non-negligible temperatures
these higher energy states would be important.
Indeed the characteristic linear in temperature
resistivity of the normal state of high
temperature superconductors persists up to
$T > 500{\rm K}$.
But we are a long way from formulating
transport theory in terms of the random
matrix approach. One must go far beyond
simple level statistics in order to
consider
the response functions of an
interacting Fermion system.

\acknowledgements
The authors wish to thank J. Bellisard, T. Dombre,
	B. Dou\c{c}ot, L. Levy, D. Poilblanc, S. Shastry, and C. Sire.


\figure{
	Level spacing distribution calculated
	using random symmetric matrices
	(with Gaussian distributed elements)
	and random diagonal matrices
	of the sizes encountered in the 14-site chain.
	Solid curve: Poisson distribution,
	Long dashed curve: Wigner surmise,
	Diamonds: Random diagonal matrices,
	Plus signs: Random symmetric matrices.
\label{RMT}
	}
\figure{
	Level spacing distribution for the 12-site
	and 14-site
	S=1/2 antiferromagnetic chains
	with open boundary conditions.
	Diamonds: L=12, Plus signs: L=14.
\label{BETHE}
	}
\figure{
	Level spacing distribution for two coupled chains
	each of length 7 for different interchain couplings.
	The points have been joined for clarity.
	Medium dashes: $J_{\perp}/J = 1.0$,
	Short Dashes: $J_{\perp}/J = 0.1$,
	Dash-dots: $J_{\perp}/J = 0.01$.
\label{JPERP}
	}
\figure{
	Interpolation parameter $I$ as a function of
	interchain coupling ${\rm ln}(J_{\perp}/Jd)$.
	Diamonds: S=0, Plus signs: S=2, Squares: S=4,
	Crosses: S=6, Triangles: S=8.
\label{LOGJPERP}
	}
\figure{
	Level spacing distribution for the
	next-nearest-neighbor coupled antiferromagnet
	for various couplings. Medium Dashes: $J_{2}/J = 1.0$,
	Short dashes: $J_{2}/J = 0.3$, Dash-dots: $J_{2}/J = 0.1$.
\label{J2}
	}
\figure{
	Interpolation parameter $I$ as a function
	of next-nearest-neighbor coupling ${\rm ln}(2J_{2}/Jd)$.
	Diamonds: S=1, Plus signs: S=3, Squares: S=5, Crosses: S=7.
\label{LOGJ2}
	}
\figure{
	Level spacing distribution for the $1/r^2$
	antiferromagnet.
\label{USR2}
	}

\begin{references}
\bibitem[*]{byline} Present address:
	AECL Research, Chalk River Laboratories, Chalk River,
	Ontario, Canada, K0J 1J0.
	E-mail: hsut@cu26.crl.aecl.ca
\bibitem[**]{jbyline} E-mail: dauriac@crtbt.polycnrs-gre.fr
\bibitem{WIGNER} E.P. Wigner, Conference on Neutron Physics
	by Time-of-Flight, Gatlinburg, Tennessee, Nov. 1 and 2,
	1956, {\it Oak Ridge Natl. Lab. Rept.} {\bf ORNL-2309},
	59 (1957).
\bibitem{PORTER} A good collection of early reprints on the
	subject of level spacing distributions and random
	matrix theory is found in
	C.E. Porter, {\it Statistical Theories of Spectra:
	Fluctuations}, (Academic Press, New York, 1965).
\bibitem{BOHIGAS}
	A recent review of random matrices and their relation
	to chaotic dynamics is
	O. Bohigas, in {\it Chaos and quantum physics,
	Proceedings of the Les Houches summer school}, edited by
	M.J. Giannoni, A. Voros, and J. Zinn-Justin (North-Holland,
 	Amsterdam, 1991).
\bibitem{WIGNERB} E.P. Wigner in {\it Can. Math. Congr. Proc.},
	(Univ. of Toronto Press, Toronto, 1957).
\bibitem{WISHART} J. Wishart, Biometrika, {\bf 20}, 32 (1928).
\bibitem{BERRY} M.V. Berry and M. Tabor, Proc. R. Soc. Lond. A
	{\bf 356}, 375 (1977).
\bibitem{CASATI} G. Casati, F. Izrailev and L. Molinari,
	J. Phys. A {\bf 24},
	4755 (1991) and references therein.
\bibitem{MONT}
	G. Montambaux, D. Poilblanc, J. Bellissard, and
	C Sire, preprint.
\bibitem{PWA} P.W. Anderson, Science, {\bf 256}, 1526 (1992).
\bibitem{RAMMAL} R. Rammal, private communication.
\bibitem{SAKAI} T. Sakai and M. Takahashi, J. Phys. Soc. Jap. {\bf 58},
	3131 (1989).
\bibitem{SHKLOVSKII} B.I. Shklovskii, B. Shapiro, B.R. Sears, P.
	Lambrianides, and H.B. Shore, preprint TPI-MINN-92/54-T.
\bibitem{JULLIEN} R. Jullien and F.D.M. Haldane, Bull. Am. Phys. Soc.
	{\bf 28}, 344 (1983).
\bibitem{MAJUMDAR}
	C.K. Majumdar J. Phys. C {\bf 3}, 911 (1969); C.K. Majumdar and
	D.K. Ghosh, J. Math. Phys. {\bf 10}, 1388 (1969);
	C.K. Majumdar {\it et al.} J. Phys. C {\bf 5}, 2896 (1972).
\bibitem{HALDANE}
	F.D.M. Haldane, Phys. Rev. Lett. {\bf 60}, 635 (1988); B.S.
	Shastry, Phys. Rev. Lett. {\bf 60}, 639 (1988); V.I. Inozemtsev,
	J. Stat. Phys. {\bf 59}, 1143 (1990); F.D.M. Haldane,
	Phys. Rev. Lett. {\bf 66}, 1529 (1991); B.S. Shastry,
	Phys. Rev. Lett. {\bf 69}, 164 (1992).
\bibitem{HALDANEA} F.D.M. Haldane, Z.N.C. Ha, J.C. Talstra, D. Bernard,
	and V. Pasquier, Phys. Rev. Lett. {\bf 69}, 2021 (1992).
\end{references}
\end{document}